\documentclass[aps,preprint,superscriptaddress,]{revtex4-1}

\usepackage{graphicx}
\usepackage{lipsum}
\usepackage{hyperref}
\hypersetup{
setpagesize=false,
bookmarksnumbered=true,%
bookmarksopen=true,%
colorlinks=true,%
linkcolor=blue,
citecolor=blue,
urlcolor=blue
}
\usepackage{easyReview}
\usepackage{color}
\newcommand {\bem}{$b_\mathrm{em}$}  
\newcommand {\ABO}{$A_2B_2$O$_7$}
  
\newcommand {\BRO}{Bi$_2$Rh$_2$O$_7$}
\newcommand {\DTO}{Dy$_2$Ti$_2$O$_7$}

\begin{document}

\title{Topological Hall effect in a non-magnetic metal interfaced to a canted antiferromagnetic insulator in perovskite oxide heterostructures}
\author{Takahiro C. Fujita}
\email[corresponding author: ]{fujita@ap.t.u-tokyo.ac.jp}
\author{Koki Omura}
\affiliation{Department of Applied Physics and Quantum-Phase Electronics Center (QPEC),
  University of Tokyo, 113-8656, Tokyo, Japan}

\author{Masashi Kawasaki}
\affiliation{Department of Applied Physics and Quantum-Phase Electronics Center (QPEC),
  University of Tokyo, 113-8656, Tokyo, Japan}
\affiliation{Center for Emergent Matter Science (CEMS), RIKEN, 351-0198, Saitama, Japan}

\begin{abstract}
  We report interfacial transport properties in \textit{in-situ} grown orthorhombic perovskite oxide heterostructures consisting of an antiferromagnetic insulator DyFeO$_3$ and a paramagnetic conductor CaRuO$_3$.
  We observe Hall effect with a step-like increase amounting to an effective magnetic field of 30~T at 20~K.
  We provide a plausible explanation in the context of topological Hall effect originating from a non-coplanar spin texture and resultant emergent field in DyFeO$_3$ associated with the scalar spin chirality.
  Our results demonstrate that the proximity effect of the emergent field at heterointerfaces is a universal physical phenomenon, while it has been reported originally in a heterointerface composed of pyrochlore oxides.
  This will greatly expand the choice of materials to the heterointerfaces for the research in emergent transport phenomena, which has been limited to single compounds with both metallic properties and special spin textures.
  Additionally, this will pave the way for possible device application of the emergent field by designing and combining perovskite oxides with versatile functionalities such as multiferroicity.
\end{abstract}

\maketitle 

Heterostructures comprised of a non-magnetic metal and a magnetic insulator have been gathering considerable attention both in fundamental and applied physics.
Although the conduction electrons in the non-magnetic metal layer do not apparently penetrate into the magnetic insulator layer, the electrical properties of the heterostructure reflect the direction of macroscopic magnetization in the insulating layer due to the proximity effect, enabling the detection of the magnetic state through electrical transport measurements~\cite{nakayama_spin_2013,chen_theory_2016,zutic-2019-ProximitizedMaterials}.

\begin{figure*}[h]
  \includegraphics[width=15.5cm]{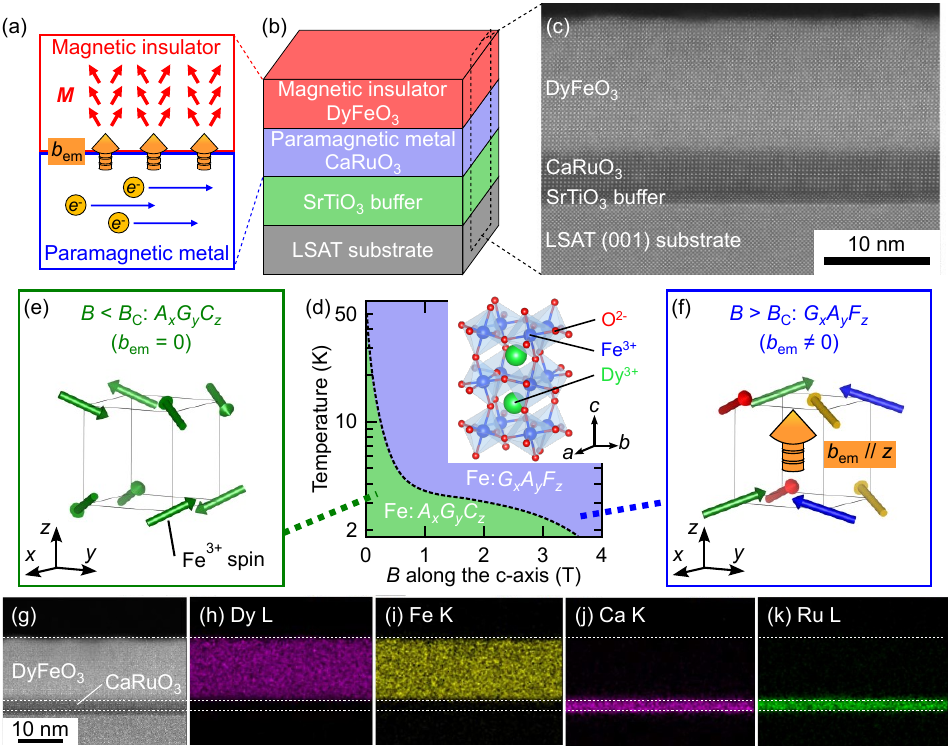}
  \caption{\label{concept}
  (a)~A schematic of interfacial emergent transport phenomena.
  Conducting electrons in a paramagnetic metal layer exhibits transport properties reflecting the emergent field ({\bem}) at the interface from a magnetic insulating layer.
  (b)~A schematic and (c)~a cross sectional transmission electron microscope (TEM) image of a DyFeO$_3$ (15 nm)/CaRuO$_3$ (4~nm) heterostructure.
  (d)~A magnetic phase diagram of Fe$^{3+}$ spin in DyFeO$_3$ as a function of temperature and magnetic field (\textit{B}) along the \textit{c} axis of DyFeO$_3$ (replotted from Ref.~\cite{wang_simultaneous_2016}).
  The inset shows the crystal structure of DyFeO$_3$.
  (e)~$A_{x}G_{y}C_{z}$ and (f) $G_{x}A_{y}F_{z}$ orders of Fe$^{3+}$ spins in DyFeO$_3$ under weak and strong \textit{B} along the \textit{c} axis, respectively.
  (g) High resolution high-angle annular dark-field (HAADF) scanning transmission electron microscopy (STEM) image of the heterostructure.
  The corresponding energy dispersive x-ray (EDX) spectrometry maps for (h) Dy L, (i) Fe K, (j) Ca K, and (k) Ru L edges.}
\end{figure*}

Very recently, we have reported a fundamentally different interfacial transport phenomenon indicating a proximity effect of the “emergent field ({\bem})” from non-coplanar spin textures (Fig.~\ref{concept}(a)) in an epitaxially grown pyrochlore oxide ({\ABO}) heterostructure consisting of non-magnetic metal {\BRO} and spin ice insulator {\DTO}~\cite{ohno_proximity_2024}.
Non-coplanar spin textures in the latter generate a scalar spin chirality acting as a fictitious magnetic field ({\bem}) that affects the transport properties in the former, resulting in the observation of the topological Hall effect (THE)~\cite{nagaosa_anomalous_2006,nagaosa_anomalous_2010}.
The observed THE in the {\BRO}/{\DTO} heterointerface suggests that the conduction electrons in the {\BRO} layer reflect not only the \textit{macroscopic} magnetization but also the \textit{microscopic} spin texture~\cite{ohno_proximity_2024}, which will open a new avenue for device applications of {\bem}.
To push such an intriguing physical phenomenon towards device functionalities, the reported heterointerface of pyrochlore oxides has a serious disadvantage in the thin film fabrication process involving unavoidable \textit{ex-situ} annealing~\cite{ohno_impact_2023}.
In this sense, replacing pyrochlore with perovskite oxides ($AB$O$_3$) will endow the interfacial THE with further controllabilities and tunabilities due to their rich knowledge in thin film research, versatility in materials choice, and capability of \textit{in-situ} growth of well-defined heterointerfaces~\cite{hwang_emergent_2012,huang_interface_2018}.

Here, we focus on interfacial magnetotransport properties in the orthorhombic perovskite oxide heterostructures consisting of DyFeO$_3$/CaRuO$_3$ (DFO/CRO), where DFO is an antiferromagnetic insulator and CRO is a paramagnetic conductor (Figs.~\ref{concept}(b) and \ref{concept}(c))~\cite{shepard_magnetic_1996,cao_itinerant--localized_1996}.
Orthorhombic DFO possesses three principal crystallographic axes (\textit{a} = 5.30~\AA, \textit{b} = 5.60~\AA, \textit{c} = 7.62~\AA)~\cite{eibschutz_lattice_1965} and undergoes unique magnetic transitions upon cooling.
In DFO, Fe$^{3+}$ spins develop a canted antiferromagnetic order with a weak ferromagnetic component along the \textit{c} axis due to the Dzyaloshinskii–Moriya interaction below $T_\mathrm{N}$ = 645~K, and exhibit a characteristic spin-reorientation (Morin) transition at $T_\mathrm{M}\approx50$~K in zero-field as shown in Fig.~\ref{concept}(d).
During this Morin transition, the configuration of Fe$^{3+}$ spins transforms from high-temperature $G_{x}A_{y}F_{z}$ (blue) to low-temperature $A_{x}G_{y}C_{z}$ (green) structures in the so-called Bertaut’s notation~\cite{bousquet_non-collinear_2016} as shown in Figs.~\ref{concept}(d)--~\ref{concept}(f).

In high-temperature $G_{x}A_{y}F_{z}$ configuration, \textit{G}-type component points toward the \textit{a} axis ($G_{x}$) but with a canting angle away from the \textit{a} axis, leading to \textit{A}-type antiferromagnetic and weak ferromagnetic components along the \textit{b} ($A_{y}$) and \textit{c} ($F_{z}$) axes, respectively, which corresponds to a non-coplanar spin structure that generates {\bem} along the \textit{c} axis.
On the other hand, low-temperature $A_{x}G_{y}C_{z}$ configuration does not yield {\bem}.
These two configurations can be switched through a magnetic field (\textit{B}) along the orthorhombic \textit{c} axis below Morin temperature $T_\mathrm{M}$ (Figs.~\ref{concept}(e) and \ref{concept}(f))~\cite{johnson_field_1980,prelorendjo_spin_1980,tokunaga_magnetic-field-induced_2008,zhao_ground_2014,ke_low_2015,wang_simultaneous_2016}, where high-temperature $G_{x}A_{y}F_{z}$ structure with finite {\bem} revives at high field (blue region in Fig.~\ref{concept}(d)).
In this study, we have detected the transition between these two magnetic states in DFO layer through the magnetotransport properties in CRO layer.
Clear Hall signal is detected due to the proximity effect of {\bem}  as a step-like structure in the Hall resistance during magnetic field scan.
At the same time, inflection points are observed in longitudinal magnetoresistance, suggesting a magnetic transition.
These anomalies in magnetotransport properties are discussed in relation to the thickness of the CRO layer, demonstrating that the observed emergent transport phenomena originate at the heterointerface.
These results indicate that the key commonality lies in non-coplanar spin textures at the heterointerface for the proximity effect of the emergent field regardless of the differences in magnetic properties and types of materials: a non-magnetic metal ({\BRO})/spin ice insulator ({\DTO}) heterostructure in previous work~\cite{ohno_proximity_2024}, an antiferromagnetic insulator (DyFeO$_3$)/paramagnetic conductor (CaRuO$_3$) heterostructure in present work.

DFO/CRO heterostructures were grown on LSAT ((LaAlO$_3$)$_{0.3}$(Sr$_2$AlTaO$_6$)$_{0.7}$) (001) substrates by pulsed laser deposition.
The thickness of CRO was varied from 4~nm to 10~nm while the thickness of DFO was set at 15~nm.
Prior to the growth of CRO, a thin SrTiO$_3$ buffer layer ($\sim$1~nm) was deposited on the LSAT substrate at 670$^\circ$C and 10$^{-6}$ Torr O$_2$.
The insertion of this buffer layer dramatically improves the surface morphology of the CRO layer (Supplementary Fig.~S1), which is a prerequisite for obtaining a sharp and well-defined heterointerface with the DFO layer.
CRO and DFO layers were deposited at 670$^\circ$C and $2\times10^{-2}$~Torr O$_2$.
To suppress oxygen deficiencies, the films were subsequently annealed \textit{in-situ} for 30 mins at 600$^\circ$C in 100 Torr O$_2$ before cooling down in the same O$_2$ atmosphere.
KrF excimer laser ($\lambda = 248$~nm) with a repetition of 5~Hz and a fluence of 2~J/cm$^2$  was employed to ablate the targets.

As confirmed in the transmission electron microscopy images and the energy dispersive x-ray spectroscopy mappings (Figs.~\ref{concept}(g)--~\ref{concept}(k)), the heterostructure is constructed as designed with indiscernible interdiffusion at the interfaces.
However, we found that the DFO layer consists of domains having their \textit{c} axes aligned either normal or parallel to the interface plane.
There can be three possible orientations when an orthorhombic perovskite grows on a cubic perovskite substrate (Supplementary Fig.~S2).
We performed electron diffraction measurements for the DFO/CRO heterostructure with 4-nm-thick CRO layer.
The DFO layer exhibits characteristic diffraction patterns with different orientations depending on the measurement points, confirming that the DFO layer has mixed crystallographic domains.
We note that the THE in CRO due to the proximity effect of {\bem} from DFO should emerge only for the domain with its \textit{c} axis normal to the interface.

\begin{figure}[h]
\includegraphics{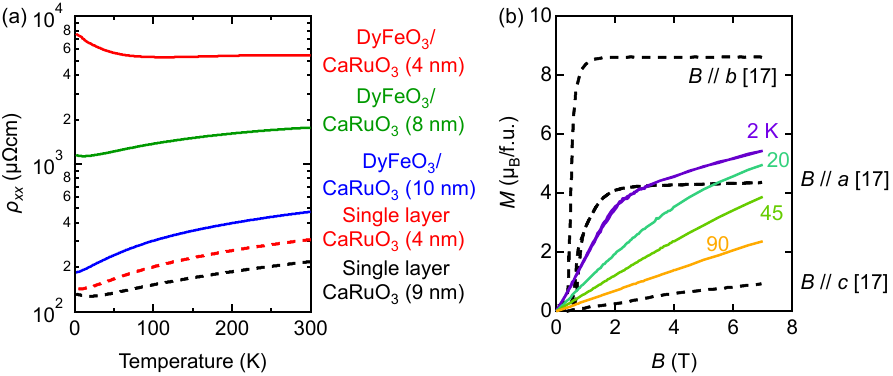}%
\caption{
(a)~Temperature dependence of the longitudinal resistivities ($\rho_{xx}$) for the three DyFeO$_3$/CaRuO$_3$ heterostructures (solid lines) and two CaRuO$_3$ single-layer films (dashed lines) with various CaRuO$_3$ thicknesses.
(b)~Magnetic field (\textit{B}) dependence of out-of-plane magnetization (\textit{M}) for a 12-nm-thick DyFeO$_3$ single layer film measured at several temperatures.
Dotted lines are anisotropic magnetization curves at 2~K for a DyFeO$_3$ single crystal along the three orthorhombic crystallographic axes taken from Ref.~\cite{zhao_ground_2014}.
\label{longitudinal}}
\end{figure}

The temperature dependence of the longitudinal resistivity $\rho_{xx}$ is shown in Fig.~\ref{longitudinal}(a) for the DFO/CRO heterostructures as well as CRO single-layer films for comparison.
The CRO single-layer films exhibit metallic conduction down to 2~K with a slight upturn below 10~K, which is comparable to those in previous reports~\cite{sakoda_transition_2023}.
The 4-nm-thick CRO shows slightly higher $\rho_{xx}$ than the 9-nm-thick film.
In the heterostructures, however, $\rho_{xx}$ at 2~K increases from $\sim$200~$\mu\Omega$cm (10~nm) to $\sim$8~m$\Omega$cm (4~nm), and the temperature dependence of $\rho_{xx}$ changes from metallic to semiconducting, indicating significant modification by the proximity effect from DFO.
As confirmed by ordinary Hall effect measurements, this modification in longitudinal transport properties is mainly attributed to the reduction in carrier mobility rather than carrier density with decreasing the CRO thickness (Supplementary Fig.~S3) plausibly originating from the enhanced interfacial scattering.
However, we cannot exclude the possibility that the additional DFO deposition process may alter the bulk properties of CRO layer in the heterostructures.
Nonetheless, even DFO/CRO (4~nm) remains conducting down to the lowest measurement temperature in this study, which enables further magnetotransport measurements reflecting the magnetic structure of DFO.

To elaborate on the magnetic properties of DFO, we performed magnetization (\textit{M}) measurements for a 12-nm-thick DFO single-layer film under \textit{B} perpendicular to the film surface with a superconducting quantum interference device magnetometer (MPMS3, Quantum Design Co.) down to 2~K and up to 7~T.
It has been reported that DFO single crystals exhibit anisotropic magnetization curves depending on the field direction along the orthorhombic axes~\cite{tokunaga_magnetic-field-induced_2008,zhao_ground_2014,ke_low_2015,wang_simultaneous_2016}.
As shown with broken lines in Fig.~\ref{longitudinal}(b), \textit{M} along the \textit{a} and \textit{b} axes at 2~K are one order magnitude larger than that along the \textit{c} axis~\cite{zhao_ground_2014}.
This difference is attributed to the antiferromagnetic ordering of Dy$^{3+}$ moments which emerges at $\sim$4~K and confines the moments in the \textit{ab} plane.
Our thin film exhibits an obvious deviation from any of the single crystal data and no saturated magnetization even at \textit{T} = 2~K and \textit{B} = 7~T.
These behaviors can be viewed as the superposition of the magnetization curves measured along the three axes for the DFO single crystals, which is not inconsistent with the fact that mixed crystallographic domain structure is confirmed in electron diffraction inspection of our thin film.

\begin{figure}[h]
\includegraphics[width=16.5cm]{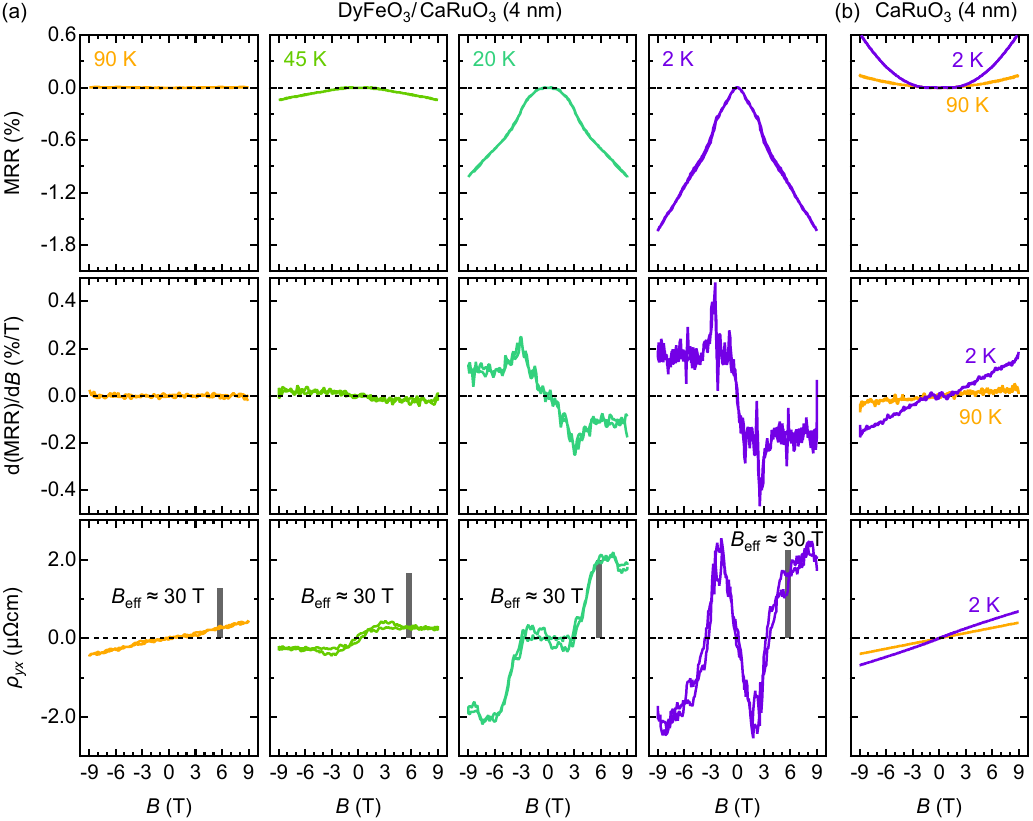}%
\caption{
(a)~Magnetic field (\textit{B}) dependence of magnetoresistance ratio (MRR), first order \textit{B} derivative of MRR (d(MRR)/d\textit{B}), and Hall resistivity ($\rho_{yx}$) for the DyFeO$_3$/CaRuO$_3$ (4~nm) heterostructure.
Note that $\rho_{yx}$ in the bottom panels is raw data deduced only by the conventional anti-symmetrization procedure, and thus contains ordinary and anomalous Hall terms.
The scale bars in the bottom panels indicate the effective magnetic field ($B_\mathrm{eff}$), which is calculated based on the Hall coefficient of the 4-nm-thick CaRuO$_3$ single-layer film.
See the main text for the definition of $B_\mathrm{eff}$.
(b)~The same data set for the 4-nm-thick CaRuO$_3$ single-layer film.
}
\label{MagnetoTransports}
\end{figure}

Figure~\ref{MagnetoTransports}(a) summarizes the magnetotransport properties of DFO/CRO (4~nm) at selected temperatures.
At 90~K, the magnetoresistance ratio (MRR $\equiv\rho_{xx}$(\textit{B})/$\rho_{xx}(0) - 1$) is almost zero, and Hall resistivity $\rho_{yx}$ is merely linear to \textit{B}.
With lowering the temperatures, negative MRR emerges and $\rho_{yx}$ becomes non-linear, suggesting the magnetic proximity effect from DFO.
MRR at 20~K exhibits an inflection point at $\sim$3~T that is visible more clearly in the \textit{B}  derivative (d(MRR)/d\textit{B}).
Interestingly at the same field, $\rho_{yx}$ drastically increases, giving rise to a step-like structure.
Upon further cooling, the inflection point in MRR becomes more prominent and the step-like $\rho_{yx}$  gradually transforms into an “N”-shape, which is plausibly related to the additional ordering of Dy$^{3+}$ spins~\cite{white_review_1969}.
These anomalous behaviors simultaneously occurred both in $\rho_{xx}$ and $\rho_{yx}$ strongly suggest a coupling with the change in the spin ordering structures from $A_{x}G_{y}C_{z}$ to $G_{x}A_{y}F_{z}$ in the DFO layer.

It is worth emphasizing that $\rho_{yx}$ presented here is raw data deduced only by the conventional anti-symmetrization procedure, and thus contains ordinary and anomalous Hall terms.
The latter term is thought to be composed of two terms: a conventional term proportional to magnetization \textit{M} and the topological term proportional to {\bem}.
It is indeed not straightforward to deconvolute these three terms in the present case because \textit{M} does not saturate within the scanned magnetic field range as shown in Fig.~\ref{longitudinal}(b).
Nonetheless, if we assume the step-like Hall signal at 20~K is ascribed solely to the topological Hall effect, the estimated effective magnetic field ($B_\mathrm{eff}$) acting on the conduction electrons reaches $\sim$30~T, with the relationship $\rho_{yx} = R_{0}B_\mathrm{eff}$ using the values $R_{0}=7\times10^{-4}$~cm$^{3}$C$^{-1}$ and $\rho_{yx}=2$~$\mu\Omega$cm.
Here, we employ the value of $R_0$ at 20~K for the 4-nm-thick CRO single-layer film, taking into account the coincidence in $R_0$ or carrier density above 90~K (red open and closed symbols in Supplementary Fig.~S3).
This procedure to estimate $B_\mathrm{eff}$ has widely been performed with the relationship $\rho_\mathrm{THE} = R_{0}B_\mathrm{eff}$, where $\rho_\mathrm{THE}$ is topological Hall resistivity.
Due to the difficulty in extracting pure $\rho_\mathrm{THE}$ component in our system, we employ $\rho_{yx}$ instead of $\rho_\mathrm{THE}$.
The carrier density of CRO in the heterostructure may differ from that of single-layer film as already discussed.
This in turn causes additional ambiguity of the estimated $B_\mathrm{eff}$, probably by a factor of 2--3 because the carrier density for the samples in this work is scattered in the same level (Supplementary Fig.~S3(a)).
On the contrary, as shown in Fig.~\ref{MagnetoTransports}(b), the CRO single-layer film shows positive MRR with quadratic \textit{B} dependence and \textit{B} linear $\rho_{yx}$ down to 2~K.
This stark contrast suggests a magnetic proximity effect from the adjacent DFO layer which modifies the magnetotransport properties.

\begin{figure}[h]
\includegraphics{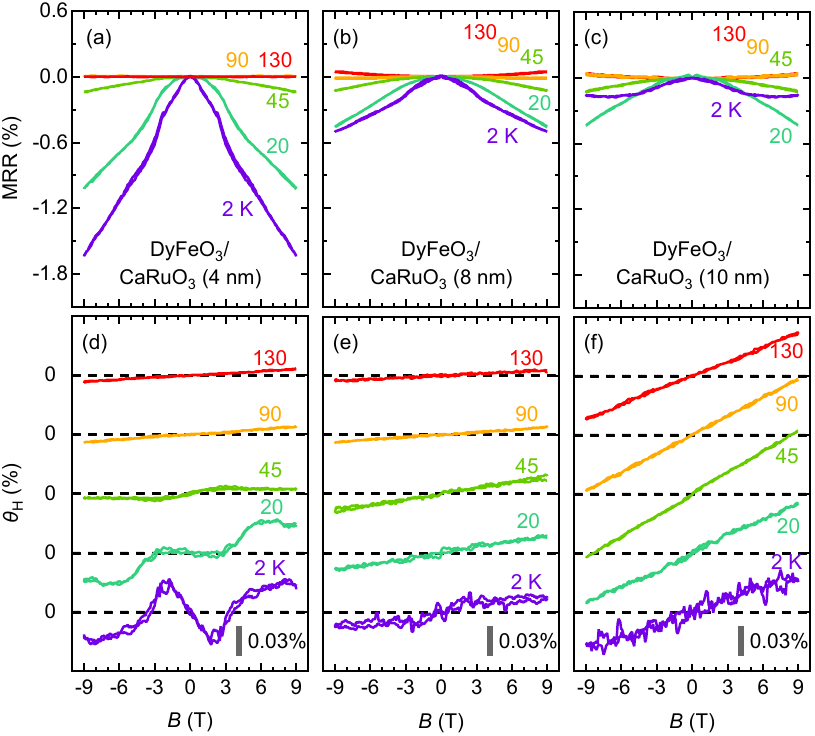}%
\caption{
Temperature dependence of magnetoresistance ratio (MRR, a--c) and Hall angle ($\theta_\mathrm{H}\equiv\rho_{yx}(B)/\rho_{xx}(B)$, d--f) for three DyFeO$_3$/CaRuO$_3$ with the CaRuO$_3$ layer thickness of (a,d)~4, (b,e)~8, and (c,f)~10 nm.
As in Fig.~\ref{MagnetoTransports}, raw $\rho_{yx}$ is used for calculating $\theta_\mathrm{H}$, and thus contains ordinary and anomalous Hall contributions.
The data in the panels (d)--(f) are vertically shifted by 0.06\% for clarity, in which the scale bars indicating 0.03\% are also shown.}
\label{ThicknessMRR_HallAngle}
\end{figure}

To corroborate the magnetic proximity effect from DFO, we show CRO thickness dependence of the magnetotransport properties in Fig.~\ref{ThicknessMRR_HallAngle}.
With increasing the thickness of CRO, the negative MR observed at temperatures lower than 45~K is gradually suppressed as seen in Figs.~\ref{ThicknessMRR_HallAngle}(a)--\ref{ThicknessMRR_HallAngle}(c).
The inflection points in MRR observed at 2 and 20~K for DFO/CRO (4~nm) (Fig.~\ref{ThicknessMRR_HallAngle}(a)) are less discernible for the other two samples with thicker CRO layer, while it is still visible at 2~K for DFO/CRO (8~nm) (Fig.~\ref{ThicknessMRR_HallAngle}(b)).
For DFO/CRO (10~nm), negative MRR becomes positive at higher magnetic fields at 2~K (Fig.~\ref{ThicknessMRR_HallAngle}(c)), indicating that the observed MRR is the superposition of the negative one from the interfacial part with the proximity effect and the positive one from the intrinsic CRO away from the interface as already shown in Fig~\ref{MagnetoTransports}(b).

The CRO thickness dependence of the Hall angle ($\theta_\mathrm{H}\equiv\rho_{yx}(B)/\rho_{xx}(B)$) exhibits consistent trends as shown in Figs.~\ref{ThicknessMRR_HallAngle}(d)--\ref{ThicknessMRR_HallAngle}(f).
Here we note again that this $\theta_\mathrm{H}$ includes all the Hall components: ordinary Hall, conventional anomalous Hall proportional to \textit{M}, and topological Hall terms.
Therefore, $\theta_\mathrm{H}$  for DFO/CRO (8~nm) and DFO/CRO (10~nm) are almost dominated by \textit{B}-linear ordinary term.
Similar to MRR, the peculiar behaviors observed in $\rho_{yx}$ for DFO/CRO (4~nm) almost diminish for the other heterostructures.
These results suggest that the proximity effect in the present DFO/CRO heterointerface decays within a few nm length scale.
This decay length, in principle, could be extracted from the thickness dependence of the transport data by assuming some decay functions.
However, quantitative analysis is quite challenging because the electron current distribution within the CRO conducting layer should vary depending on the CRO thickness.
To address this issue, we would require more systematic and precise control of CRO thickness especially in the thinner region that goes beyond the scope of this paper.

In summary, we have investigated the interfacial magnetotransport properties for DyFeO$_3$/CaRuO$_3$ heterostructures, where the emergent field from DyFeO$_3$ layer can be switched through the magnetic transition.
Above all, we have observed the topological Hall effect plausibly due to the non-coplanar spin texture in the magnetic field-induced canted antiferromagnetic state in DyFeO$_3$, yielding an effective magnetic field as large as $\sim$30~T at 20~K.
Accordingly, we find another example of the proximity effect of the emergent field at an epitaxial oxide heterointerface, demonstrating the universality of this phenomenon observed originally in pyrochlore oxide heterointerface. 
The common feature is that the emergent field from the magnetic insulating layer penetrates through the heterointerface and affects the electron transport in the conducting layer, in spite of such differences of crystal structures, spin structures generating scaler spin chirality, and elements hosting localized spins therein.
This provides unique opportunities enabled only at the heterostructures to explore physical effects and functionalities that are impossible in single bulk materials, such as non-reciprocal transports originating from the interfacial emergent field, and electric field control of magnetic structure to modulate the emergent field in multiferroic insulators.
In this regard, orthoferrites including DyFeO$_3$ employed in this work will be of great interest for realizing device application of the emergent field as prototypical multiferroic materials. 
At the same time, this class of compounds has recently been gathering considerable attention in the context of altermagnetism~\cite{naka_perovskite_2021,naka_anomalous_2022}.
Heterointerface structure will serve as a platform to test the unique properties expected for this class of magnetism, such as spin current generation and anomalous Hall effect.

\section*{supplementary material}
See the supplementary material for the buﬀer layer eﬀect on surface morphology (Fig.~S1), the electron diffraction measurements (Fig.~S2),  and temperature dependence of carrier density and mobility of the samples (Fig.~S3).

\section*{acknowledgements}
This work was partly supported by JSPS Grants-in-Aid for Scientific Research (S) No. JP22H04958, JSPS Grant-in-Aid for Early-Career Scientists No. JP20K1516, The Mitsubishi Foundation, Kazuchika Okura Memorial Foundation, Yazaki Memorial Foundation for Science and Technology, and The Tokuyama Science Foundation.
The magnetic measurements were performed using the facilities of the Cryogenic Research Center, the University of Tokyo.

\section*{Conflict of Interest}
The authors have no conflicts to disclose.

\section*{Data Availability Statement}
The data that support the findings of this study are available from the corresponding author upon reasonable request.

\bibliography{DFOCRO}

\newpage
\section*{Figures}
\textbf{}

\newpage
\textbf{}
\newline
\newline
\newline
\newline
\newline
\newline

\newpage
\textbf{}
\newline

\newpage
\textbf{}
\newline
\newline
\newline
\newline
\newline

\end{document}